\documentclass{mem}
\usepackage{natbib}\usepackage{txfonts}\usepackage{balance}
\usepackage{graphicx}
\usepackage[a4paper]{hyperref}
\idline{75}{282}
\begin{document}
\def\teff{$T\rm_{eff }$}
\def\kms{$\mathrm {km s}^{-1}$}
\newcommand{\lsp}{LS~I~+61$^{\circ}$303}
\newcommand{\lsi}{LS~I~+61$^{\circ}$303~}
\newcommand{\grs}{GRS 1915+105~}

\title{Astrophysical Jets : what can we learn from Solar Ejections?}

   \subtitle{}

\author{M. \,Massi\inst{1}
\and G. \,Poletto\inst{2} 
          }

  \offprints{M. Massi}

\institute{
Max Planck Institute for Radio Astronomy,
Auf dem Huegel 69,
D-53121 Bonn,
Germany
\email{mmassi@mpifr-bonn.mpg.de}
\and
INAF -- Arcetri Astrophysical Observatory,  
Largo Fermi 5,                     
I-50125 Firenze,                      
Italy                              
\email{poletto@arcetri.astro.it}
}

\authorrunning{Massi \& Poletto}

\titlerunning{Astrophysical Jets and Solar Ejections}

\abstract{Ejections from the Sun can be observed with a higher resolution than 
in any other astrophysical object: can we build up on solar results
and apply them to astrophysical objects?   
Aim of this work is to establish whether there is any analogy between 
solar ejections and ejections in microquasars and AGNs. Briefly reviewing 
jets properties from these objects and from the Sun, we point out some
characteristics they share  and indicate research areas where cross-breeeding
between astrophysical and solar research is likely to be productive.
Preliminary results of this study suggest, for instance, that there may be an 
analogy between blobs created by tearing instability in  
current sheets (CSs) associated with solar coronal mass ejections (CMEs) 
and  quasi periodic ejections of plasma  associated with large  
radio outbursts in microquasars.
}
\maketitle{}

\section{Introduction}

Radio and X-ray observations of microquasars show the occurrence of
two types of jets with completely different characteristics. 
The first type, referred to as steady jet, is a quasi-steady slowly moving
continuous ejection emerging from an optically thick (flat or inverted radio 
spectrum) radio core region. 
The second type of jet, referred to as transient jet, 
shows up with an optically thin radio outburst
and features a sequence of 
bright regions 
moving at superluminal speeds 
\citep{fender04, massi10}. 
The formation of steady jets is recognized to be mediated by 
large-scale open magnetic fields threading the rapidly rotating accretion 
disks around compact objects (i.e. jets are accelerated by magneto-centrifugal 
forces, \citet{blandford82}). 
The transient radio jets,
are instead related to  shocks
\citep{marscher85,turler10}.
These shocks 
 seem to be  caused by relativistic plasma travelling 
in the slowly moving pre-existing steady jet, established during a  previous 
phase.
In other words the two jets seem to be  inter-related \citep{fender04}.
How does the steady-to-transient jet transition occur? A possible mechanism 
triggering the switch could be magnetic reconnection.
In analogy with solar flares, magnetic energy is probably built-up and
accumulated over long time scales during the steady jet phase 
and then dissipated over very short time 
scales in explosive events \citep{komissarov06}.
Such events occurring in the central engine, close to the black hole and 
accretion disk, can be the source of ejections/flaring events possibly causing shocks
\citep{massi10}.
Solar ejections have been observed with a higher resolution than attainable in 
any other
astrophysical objects. Are there analogies between ejections in different
objects?
In the following we briefly illustrate microquasar and AGN ejections 
(Sec. 2) and solar ejections (Sec. 3) highlighting properties they may share. 
Our conclusions are summarized in the last Section (Sec. 4).
\section{Microquasars and AGNs}
\subsection{Quiet and active coronas in accretion disks and the Sun}
Numerical simulations  show that  
during the {\it low/hard} X-ray state, corresponding to the radio steady jet,
a corona, similar to the solar corona,
is present around the accretion disk. 
Starting with a differentially rotating torus
threaded by toroidal magnetic fields,
the simulations show that magnetic flux 
buoyantly escapes from the disk and creates loop like structures similar
to the solar coronal loops (\citet{machida00},  \citet{yuan09} and references there).
As for the solar corona, magnetic fields are
likely to play a fundamental role, although, as shown by \citet{zhang00}, 
there is a factor 500 between a) the magnetic fields of the inner regions 
of an accretion disk around a stellar mass 
black hole and the active regions of the Sun and, b),
the temperatures of the accretion disk coronae and the solar corona.
The accretion disk corona is likely to be heated  from small-scale reconnection
events, in a scenario analogous to the microflare heated solar corona suggested
by e.g. \citet{parker63,benz98,krucker00}.
Also, large coronal loops can emerge at the surface of the differentially 
rotating accretion disc where, because of the ensuing shear of 
the large magnetic structure, more energetic reconnection phenomena may be triggered.
As far as transient ejections are concerned, \citet{yuan09} built an
MHD model for ejections  in Sgr A*, based on the analogy between 
astrophysical jets and coronal mass ejections (CMEs). 
\citet{yuan09} argue that the passage of these high speed 
plasmoids through the steady jet may indeed form shocks which show up as bright knots 
embedded in the steady continuous jet.
\subsection{Quasi periodic oscillations (QPOs) in astrophysical objects}
Quasi periodic oscillations of 20 min, observed around the supermassive 
black hole Sgr A*, modulating large NIR/X-ray flares lasting 
about 100 min, have been ascribed to an active region (AR) in the inner orbit 
of the accretion disk. More precisely, an AR within an accretion disk
seen at high inclination, is supposedly located in an orbit with a rotation 
period of 20 minutes and is able to survive for 100 minutes (since 5 cycles
have been observed), giving rise to the observed QPOs 
\citep{eckart06,eckart08}.
Can we compare the total lifetime of the QPOs in Sgr A* with the durations of the long 
lasting solar CMEs? 

Similar kind of QPOs, i.e.   related to the inner orbit  of the accretion disk,
have been observed also
in stellar mass black holes \citep{stella98}.
In these cases, 
milliseconds QPOs result because of the  smaller orbit. 
However, in three X-ray binaries 
different kind of QPOs
have been observed in radio and X-rays.
They are closely related to large radio outbursts and  have
quite  long timescales, i.e. minutes/hours.
During the decay of a radio  outburst
a variety of   22$-$120 min  oscillations were observed at radio wavelenghts in
V404 Cyg \citep{hanhjellming92}.
Radio  oscillations of 30$-$84 min  were
observed in \lsi  during the decay 
of  radio outbusts \citep{peracaula97}, whereas  
\citet{harrison00} observed oscillations of  30 min in X-rays 
during the onset of a radio outburst. 
In \grs \citet{fender99} observed at radio wavelengths
similar long-term QPOs associated with a large optically thin 
outburst.
Simultaneous QPOs observations of \grs in X-rays, infrared and radio wavelengths 
have been interpreted as periodic ejections of plasma with subsequent 
replenishment of the inner accretion disk. The estimated mass of these blobs 
ejected every few tens of minutes are on the order of $\sim 10^{19}$g (see 
references in \citet{mirabel99}). 
As to AGNs, QPOs of 55 minutes have
been associated with the flat spectrum radio quasar 3C 273 
\citep{espaillat08}
and QPOs of $~$60 minutes have been observed in the narrow-line Seyfert 
1 Galaxy RE J1034+396 \citep{gierlinski08}. 
\citet{rani10}  recently found evidence for  QPO of 15
minutes in  the blazar
S5 0716+714.
These observations  in V404 Cyg, \lsi and \grs
of  20$-$120 min  oscillations
related to  large outbursts and associated to  periodic ejections of plasma,
naturally lead us to questions like: 
Is there any 
evidence  for periodic minor solar ejections  related to  large 
CMEs? 
\section{Solar ejections}
\subsection{Coronal mass ejections (CMEs)}
Mass ejections on the Sun have been recognized recently to be far more numerous
than previoulsy predicted and their size and energy span over several order of 
magnitudes, from the mini X-ray polar jets to the spectacular  
{\it Coronal Mass Ejections (CMEs)}: 
events where the ejection of mass is unambiguously observed
as an enhanced density structure moving outwards through the solar corona and
the heliosphere. The average mass of a CME is on the order of $10^{15}$ g,
and the speed can reach up to nearly 3000 km s$^{-1}$ with an average value of
450 km s$^{-1}$. The highest kinetic energy of CMEs is $10^{32} - 10^{33}$ erg,
with an average of (2 - 3) $10^{30}$ erg. 
The lifetime of CMEs is not well defined, as it depends on the level
where they are observed:
often CMEs are associated with well identifiable solar disk phenomena, that 
last several hours, while coronal manifestations of CMEs can be observed for 
days. Heliospheric CMEs, called ICMEs (Interplanetary CME), can be followed 
over longer times. Depending on the phase of the solar activity cycle, the 
frequency of occurrence of CMEs varies between a few events/day and a few 
events/week. 
As we said, there is a whole family of ejections, in the Sun. We may ask whether
they originate from the same physical processes or whether they include 
different physical processes. Because CMEs' mass, kinetic energy 
\citep{vourlidas02}
and width (at least between 20 and 120 degrees, \citet{robbrecht09}), 
follow a power-law distribution, most likely there is a 
continuous distribution of mass ejections,
from the largest to micro-events, i.e. CMEs are scale-invariant events.
This result hints to a common physical origin for CMEs, likely to be identified
with magnetic reconnection and subsequent energy release. This raises a 
question: Can astrophysical mass ejections be accounted for by the
same mechanism?
\subsection{CME-associated current sheets}
Models of solar ejections invoke topological changes of the magnetic field:
for large CMEs the catastrophic loss of equilibrium of a magnetic flux rope
stretches the preexisting magnetic configuration creating a Current Sheet (CS) (see Fig.1).
Here we focus on the CME flux rope model of \citet{lin00}
that has been invoked to explain episodic jets in black hole systems as well 
\citep{yuan09}.
\begin{figure}
\resizebox{\hsize}{!}{\includegraphics[clip=true]{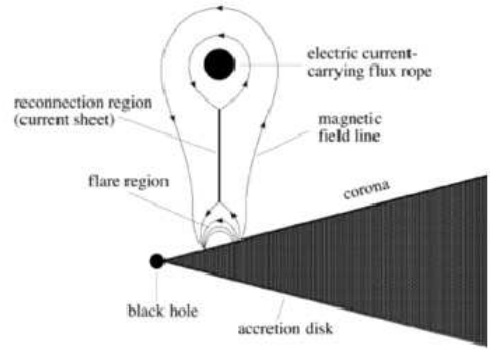}}
\caption{\footnotesize 
The CME bubble is connected 
to the low-lying loops via an extended current sheet \citep{yuan09}.}
\end{figure}
Indeed, CSs, at the model-predicted position, have been detected via
spectroscopic analyses of solar data. Also, properties of the
diffusion region around the CS (likely the CS itself is too tiny to
be identified) have been derived: 1) the high electron temperatures seen at 
the CS site slowly decays after the CME passage over hours/days; b) a high 
kinetic temperature - on the order of a few units $\times 10^7~K$ characterizes
the diffusion region, slowly decaying
and hinting to the presence of turbulence within the CS; c) {\bf densities in
the diffusion region seem to be up to one order of
magnitude higher than ambient densities} (see, e.g. \citet{poletto09,schettino10} and references
therein).
Continuing with our attempts to search for features shared by astrophysical and
solar jets we may ask: is there any possibility to identify
high temperature/density signatures associated with outward propagating 
astrophysical ejections? 
Can astrophysical blobs be interpreted in terms of reconnection?
Analysis of the temporal behavior of the CME-associated diffusion regions, in
the Sun, allowed us to detect
{\it blobs} moving outward along the CS structure. {\bf These
outflowing features can be  interpreted in terms of tearing instability within
the CS.} 
Tearing instability promotes the formation of magnetic islands, usually identified with
the above mentioned blobs. Reconnection is thus unsteady, or {\it bursty}
\citep{tanuma05}.
\section{Conclusions}
This brief analysis of astrophysical and solar jets revealed a number of
analogies among ejections events that differ by order of magnitudes in energy,
size, mass. First attempts to bridge the gap between the Sun and other objects
have focussed on applying a popular CME model to accretion disk jets. Other
so far unexplored areas include the interpretation of  minute time scale   oscillations in terms of 
reconnection-related phenomena. Altogether this research area is still in its 
infancy, but appears to deserve more attention than got so far.  

\begin{acknowledgements}
G. P. acknowledges support from ASI I/015/07/0.
\end{acknowledgements}

\bibliographystyle{aa}

\end{document}